\documentstyle[aps,preprint]{revtex}
\begin{document}
\draft
\preprint{IMSc-97/11; hep-th yy/mm/nnn}
\title{Black Hole Radiance and Supersymmetry}
\author{Parthasarathi Majumdar\footnote{email: partha@imsc.ernet.in}}
\address{The Institute of Mathematical Sciences, CIT Campus, Madras
600113,  India.}
\maketitle
\begin{abstract}
Starting with free massless scalar and spinor fields described by a globally $N=1$ 
supersymmetric action, infalling on a Schwarzschild black hole, the outgoing Hawking 
radiation is shown to break supersymmetry spontaneously, exactly as induced by a heat 
bath in Minkowski space, with no generation of Nambu-Goldstone fermions.
\end{abstract}

Bose-fermi asymmetry appears to be intrinsic to the formulation of singularity theorems 
in general relativity \cite{hawkell}. The weak energy condition is certainly not valid 
for spinor fields, while for classical bosonic particles it is perfectly consistent with  
phenomena like 
the Penrose process \cite{wald}. Perhaps for related reasons, this dissimilarity persists 
through quantization of the respective matter fields in classical black hole spacetimes. 
Indeed, superradiance of spin-1/2 particles is severely suppressed, in contrast to 
integral spin fields of appropriate charge/angular momentum \cite{unruh}. Furthermore, in 
Hawking's celebrated theory of black hole evaporation \cite{hawk}, integral spin fields are 
radiated 
in a Planckian (Bose-Einstein) spectrum at a `temperature' proportional to the surface 
gravity of the black hole. Spinor fields, on the other hand, are radiated in accord with 
Fermi-Dirac statistics at the same temperature. Of course, one can scarcely overemphasize 
that this `thermal' nature of the Hawking spectrum has {\it no} obvious statistical 
mechanical underpinning, in contrast to standard thermodynamics; it is arrived at by 
analogy, albeit a compelling one. Until satisfactory micro-foundations for this are 
discerned, signals of thermal behaviour in 
semi-classical black hole physics ought, in our opinion, to be carefully analyzed on a 
case-by-case basis. This spirit underlies the present investigation.

We focus on a situation where the infalling matter is (globally) supersymmetric to begin 
with; in 
particular, at past null infinity (${\cal I}^-$), we envisage a globally supersymmetric 
model of noninteracting massless complex scalar and chiral spinor fields. The Hilbert 
space of the 
theory has a unique supersymmetric vacuum with particle-excitations forming irreducible 
representations of the supersymmetry algebra. Now, any state on ${\cal I}^-$ will evolve 
into a state on the event horizon (${\cal H}^+$), belonging to one of two mutually 
exclusive (in absence of backreaction) classes, viz., those which are 
purely outgoing, i.e., have zero Cauchy data on ${\cal H}^+$ and support on ${\cal 
I}^+$, and those which have zero Cauchy data on ${\cal I}^+$ and support on ${\cal H}^+$. 
As is well-known \cite{hawk}, an inherent ambiguity in the latter is chiefly responsible 
for the thermalization of the radiation received at ${\cal I}^+$. It is obvious that 
this randomness emasculates the supermultiplet structure one had at ${\cal I}^-$. Our 
concern here is with the nature of the violence, especially in comparison with 
supersymmetry breakdown induced by finite temperature effects in {\it Minkowski} space 
\cite{gg}. Once again, there is no a-priori reason for these two phenomena to be 
identical, but they turn out to be so, in unexpected detail.

The scalar and spinor fields in our model have the following expansion (at ${\cal I}^-$),
\begin{eqnarray}
{\phi} ~ &=& ~\sum_k \frac {1}{\sqrt{2 \omega_k}}~\left ( a^B_k f^B_k~+~ b_k^{B ~ \dag} 
{\bar f_k^B} \right) ~\nonumber \\
{\psi_+} ~ &=& ~\sum_k \frac {1}{\sqrt{2 \omega_k}}~ \left( a^F_{k,+} 
f^F_k~+~b_{k,-}^{F~ \dag} {\bar f_k^F} \right) u_{k,+}~, \label{flxp}
\end{eqnarray}
where, the $\{ f_k \}$ are complete orthonormal sets of solutions of the respective field 
equations, with positive frequencies only at ${\cal I}^-$,  and $u_{k,+}$ is a positively 
chiral spinor, reflecting the chirality of 
$\psi_+$. The creation-annihilation operators obey the usual algebra, with $B$ ($F$) 
signifying Bose (Fermi). The conserved N\"other supersymmetry charge is given in terms of 
these creation-annihilation operators by (at ${\cal I}^-$)
\begin{equation}
Q_+({\cal I}^-)~=~\sum_k \left( a^F_{k,+} 
{b^B_k}^{\dag}~-~{b^F_{k,-}}^{\dag} a^B_k \right) u_+(k)~,\label{qu} 
\end{equation}
and annihilates the vacuum state $|0_->$ defined by
\begin{equation}
a^{B,F}_k |0_->~=~0~=~b^{B,F}_k |0_->~~. \end{equation} 

The existence of two disjoint classes of states at the horizon, as mentioned earlier, 
imply that the fields also admit the expansion \cite{hawk} 
\begin{eqnarray}
\phi ~ &=& ~ \sum_k \frac{1}{ \sqrt{2 \omega_k}} ~ \left(~ A_k^B p_k^B~+~B_k^{B~ \dag} 
{\bar p}_k^B ~+~A_k^{'B} q_k^B~+~ B_k^{'B~ \dag} {\bar q_k^B} ~ \right)~\nonumber \\
\psi_+ ~ &=& ~\sum_k \frac{1}{\sqrt{\omega_k}}~ \left(  A_{k,+}^F p_k^F~+~ B_{k,+}^{F~ \dag} 
{\bar p}_k^B 
~+~A_{k,+}^{'F} q_k^F~+~ B_{k,-}^{'F~ \dag} {\bar q_k^F} \right)~u_+(k)~~, \label{flxp+} 
\end{eqnarray}
where, $\{p_k\}$ are purely outgoing orthonormal sets of solutions of the respective 
field equations with positive frequencies at ${\cal I}^+$, while $\{q_k\}$ are orthonormal 
sets of solutions with no outgoing component. The final vacuum state $|0_+>$, defined by 
the requirement 
\begin{equation}
A^{B,F} |0_+>~=~0~=B^{B,F} |0_+>~=~A'^{B,F} |0_+>~=~B'^{B,F} |0_+>~ \end{equation} 
is not unique, because of the inherent ambiguity in defining positive frequency for the 
$\{q_k\}$; in fact, one can write $|0_+> = |0_I> |0_H>$ with the unprimed (primed) 
operators acting on $|0_I>$ ($|0_H>$). The ambiguity in $|0_H>$ results in particle 
creation at the horizon with eventual radiation out to future infinity as a thermal 
spectrum of positive energy particles.  Note, however, that a supersymmetry charge 
$Q({\cal I}^+)$ may indeed be defined, analogously to eqn. (\ref{qu}), in terms of the 
unprimed operators, and that $Q({\cal I}^+) |0_+>~=~0$. Such a charge also satisfies the 
$N=1$ superalgebra at ${\cal I}^+$. 

The field operators $a_k, b_k$ at ${\cal I}^-$ are of course related to the $A_k, B_k$ 
and $A'_k, B'_k$ through the Bogoliubov transformations
\begin{eqnarray}
A_k^B~&=&~ \sum_{k'} \left(~\alpha_{kk'}^B a_{k'}^B~+~\beta_{kk'}^B 
b_{k'}^{B ~\dag}~\right)~\nonumber\\
B_k^B~&=&~ \sum_{k'} \left(~\alpha_{kk'}^B b_{k'}^B ~+~\beta_{kk'}^B a_{k'}^{B~\dag}~ 
\right)~\nonumber \\
A_{k,+}^F~&=&~\sum_{k'} \left(~\alpha^F_{kk'} a^F_{k', +}~+~\beta^F_{kk'} b_{k', 
-}^{F~ \dag}~\right)~ \nonumber \\
B^F_{k',-}~&=&~\sum_{k'} \left(~\alpha_{kk'}^F b_{k',-}^F~+~\beta_{kk'}^F a^{F~ \dag}_{k', 
+}~\right)~, \label{bog}
\end{eqnarray}
and similarly for the primed operators. We notice in passing that  
\begin{equation}
Q({\cal I}^-) |0_+>~\neq~ 0~,~ Q({\cal  I}^+) |0_->~\neq~0~,~ for 
~\beta^{B,F}~\neq~0~.\label{hnt} \end{equation} 

The issue the we now wish to focus on is whether the radiated system of 
particles has $N=1$ spacetime supersymmetry. To address this question, recall that vacuum 
expectation values (vevs) of observables at future null infinity are defined by \cite{hawk}
\begin{equation}
<{\cal O}>~\equiv~<0_-| {\cal O} |0_->~=~Tr (\rho {\cal O})~ \label{vev}
\end{equation}
where, $\rho$ is the density operator. The trace essentially averages over the 
(nonunique) states going through the horizon, thus rendering the vevs of observables (at 
${\cal I}^+$) free of ambiguities. We also recall that a 
sufficient condition for spontaneous supersymmetry breaking is the existence of {\it a} 
fermionic operator ${\cal O}$ which, upon a supertransformation, yields an operator with 
non-vanishing vev, i.e., $< \delta_S {\cal O}>~\neq~0$. Thus, if one is able to show that 
for {\it all} fermionic observables ${\cal O}({\cal I}^+)$,
\begin{equation}
<0_-| \delta_S {\cal O}({\cal I}^+) |0_->~=~0~, \label{ss} \end{equation}
then we are guaranteed that the outgoing particles form a supermultiplet. 

However, this is not the case, as is not difficult to see; for, consider the supercharge 
operator itself at ${\cal I}^+$. Using the supersymmetry algebra, it can be shown that
\begin{equation}
<0_-|\delta_S Q({\cal I}^+)|0_->~=~{\bar \epsilon} \gamma_{\mu}~<0_-|P^{\mu} ({\cal 
I}^+)|0_->~,\label{mom} \end{equation}
where $P^{\mu}$ is the momentum operator of the theory. In our free field theory, the rhs 
of (\ref{mom}) is trivial to calculate, using eq. (\ref{bog}) above, so that we obtain
\begin{eqnarray}
<~\delta_S Q({\cal I}^+) ~>~ &=& {\bar \epsilon} \gamma~\cdot \sum_k k 
<~N^B_k~+~N^F_k~>~\nonumber \\
&=& {\bar \epsilon} \gamma ~\cdot~ \sum_{k,k'} k \left(~ 
|\beta_{kk'}^B|^2~+~|\beta^F_{kk'}|^2 ~\right)~~.\label{ssbr} 
\end{eqnarray}
Thus, supersymmetry is spontaneously broken in the sense described above, so long as the 
Bogoliubov coefficients $\beta^{B,F}$ are non-vanishing\footnote{ A hint of this was already 
available from eq. (\ref{hnt})}. In fact, we know from Hawking's seminal work \cite{hawk} 
that 
\begin{eqnarray}
<~N_k^B~>~ &=& ~ \sum_{k'} |\beta^B_{kk'}|^2~=~|t_{|k|}|^2~\left(~e^{2\pi 
|k|/\kappa}~-~1~\right)^{-1}~\nonumber \\
<~N_k^F~>~ &=&~ \sum_{k'} |\beta^F_{kk'}|^2~=~|t_{|k|}|^2~\left(~e^{2\pi|k|/ 
\kappa}~+~1~\right)^{-1}~.\label{haw}
\end{eqnarray}
Here, $\kappa$ is the surface gravity of the Schwarzschild black hole and given by 
$M/4$ where $M$ is the black hole mass. 

It is instructive to compare these results, in particular, those expressed by eq.s 
(\ref{ssbr}) and (\ref{haw}), with a flat space calculation of the thermal ensemble 
average of  $\delta_S Q$, assuming an ideal gas of massless scalar and chiral spinor 
particles in 
equilibrium with a heat bath at a temperature $T$. Following \cite{gg}, finite 
temperature field theory can be used to show that, in our theory, 
\begin{eqnarray}
<\delta_S Q>_T~ & \equiv & ~{ {\int {\cal D} \phi {\cal D} \psi e^{-S/T} \delta_S Q} 
\over {\int {\cal D} \phi {\cal D} \psi e^{-S/T}}}~\nonumber \\
&=&~ {\bar \epsilon} \gamma ~\cdot~ \sum_k k \left[~n_k^B(T)~+~n_k^F(T)~\right]~, \label{fts}
\end{eqnarray}
where, $n_k^{B,F}(T)~=~( e^{\omega_k / T}~ \mp~1)^{-1}$. With the identification between 
the temperature $T$ of the heat bath and the Hawking temperature $T_H \equiv 
\kappa/2\pi$, the similarity between supersymmetry breakdown in the two cases is 
unmistakable. 

Next consider the possible generation of Nambu-Goldstone fermions; the standard proof of 
the Goldstone theorem has to be adapted to the black hole geometry. We shall ignore 
subtleties that this adaptation might elicit and make the following seemingly reasonable 
assumptions: (a) the generally covariant definition of the time ordered product in curved 
backgrounds, given e.g., in \cite{wald}, can be written as 
\begin{equation}
{\cal T}~{\cal S}_{\mu}(x) {\cal O}(y)~=~{\cal S}_{\mu} {\cal O} \Theta (x,y)~+~{\cal 
O}{\cal S}_{\mu} \Theta(y,x)~, \label{tor} \end{equation}
where, 
\begin{eqnarray}
\Theta(x,y)~ &=&~~1~~, when~x~\in~J^+(y) \nonumber \\
             &=&~ 0~~ otherwise~; \label{thta}
\end{eqnarray}
with $J^+(y)$ being the {\it causal future} of $y$;
(b)the $\Theta$ function is also assumed to satisfy\footnote{This is probably a valid 
assumption in a static metric like Schwarzschild}
\begin{equation}
{\cal S}^{\mu} \nabla_{\mu,x} \Theta (x,y) ~=~{\cal S}^0 \delta (x^0-y^0)~
\end{equation}
(c) the orthonormal set of purely outgoing solutions $\{ p_k(x) \}$ is to serve as the 
kernel for Fourier transformations, with the assumed properties
\begin{eqnarray}
k_{\mu} p_k(x)~& = &  ~-i \nabla_{\mu} p_k(x)~\nonumber \\ 
\lim_ {{\vec k} \rightarrow 0} p_k({\vec x},0)~ & = &~1~. \label{pk}
\end{eqnarray}
One might object to the apparent lack of manifest covariance in the above, but one hopes 
that the inferences to follow would be vindicated by a more rigorous treatment.
Consider now the quantity
\begin{equation}
\lim_{{\vec k} \rightarrow 0} k^{\mu} {\cal F}_{\mu}(k)~\equiv~\lim_{{\vec k} \rightarrow 0} 
{\bar \epsilon} \int d^4 x \sqrt{-g} ~
k^{\mu}p_k(x)  <0_-| {\cal T} {\cal S}_{\mu}(x) {\cal O}_(y) |0_->~, \label{gt1} 
\end{equation}
where, $\epsilon$ is the supersymmetry parameter, ${\cal S}_{\mu}$ is the N\"other 
supercurrent and ${\cal O}$ a local spinorial operator located at $y \in {\cal I}^+$. With 
the above assumptions, it is not difficult to show that 
\begin{eqnarray}
\lim_{{\vec k} \rightarrow 0} k^{\mu} {\cal F}_{\mu}~&=&~{\bar \epsilon} <0_-| \delta_S {\cal 
O}(y)|0_-> ~\nonumber \\
&-&i \lim_{{\vec k} \rightarrow 0} \int d^4 x \nabla_{\mu} \left ( p_k(x) <0_-| 
{\cal T} {\cal S}^{\mu}(x) {\cal O}(y)|0_-> \right)~.\label{gt2}
\end{eqnarray}
The surface term in (\ref{gt2}) is usually taken to vanish, so that, if $<\delta_S {\cal 
O}(0)>~\neq~0$, a massless pole -- the Nambu-Goldstone pole appears. However, in the 
situation under consideration, 
the surface term need not vanish, mainly because of eq. (\ref{hnt}), although one does 
have $Q({\cal I}^-) |0_->=0$. Thus, it is no longer true that the rhs of (\ref{gt2}) is 
unambiguously nonzero, and hence the existence of the massless pole cannot be inferred. 

Once again, similar conclusions ensue from Minkowski space considerations at finite 
temperature: recall that conventionally, one Euclideanizes and compactifies the time 
coordinate and imposes 
suitable boundary conditions on the fields (viz., periodic or antiperiodic) \cite{dol} with 
respect to this coordinate. As can be easily shown \cite{gg}, the surface terms do not 
vanish because of contributions from the boundary of the Euclideanized 
time coordinate. In our case, of course, no such compactification of coordinates is 
necessary, but similar non-vanishing of the surface terms results because of the 
nontriviality of the Bogoliubov coefficients $\beta$, and the identification of the 
surface gravity with the (Hawking) temperature. 

The manner of supersymmetry breakdown in outgoing Hawking radiation, at least in the 
simple example dealt with here, resembles finite temperature supersymmetry breaking 
effects substantially. Indeed, from the perspective of the Euclidean functional integral 
approach \cite{hh} one may consider the foregoing results as obvious. That the same 
results emerge without appealing to the premises of such an approach is perhaps of some 
importance, especially in view of recent calculations \cite{call} of the radiation 
formula, for some special cases, using the theory of Dirichlet branes \cite{pol} and their 
agreement \cite{dasm} with results derived within the semiclassical analysis. The 
issue of supersymmetry breaking considered here will be important in those cases as well, 
more so because the Dirichlet branes under consideration are assumed to have some unbroken 
spacetime supersymmetry in most cases. 

Extensions of the results to charged or rotating black holes may be accomplished by 
incorporation of a `chemical potential': $\omega_k \rightarrow \omega_k - \mu~where~\mu= e 
\Phi ~or~m \Omega$ \cite{hawk}. It would be interesting to consider `supersymmetric' black 
holes in supergravity 
theories, although one expects them to behave similarly as long as they emit 
Hawking 
radiation. Emission of photons and gravitons and their superpartners should not change 
the results qualitatively, so long as their mutual interactions can be neglected. One 
does not expect a breakdown of full nonlinear local supersymmetry in the conventional 
sense though;
the absence of Nambu-Goldstone spinors  should inhibit the super-Higgs effect, so that 
gravitini retain their masslessness. However, this issue clearly warrants a more thorough 
analysis. 

Finally, a word on possible application of our results; clearly, Hawking radiation is 
physically irrelevant (i.e., from an observational standpoint) for large black holes 
because these typically would have very small surface gravity (Hawking temperature). The 
arena where this process becomes important is that of the early universe where density 
fluctuations may produce low mass black holes \cite{hawk}. Now supersymmetry is known to 
play an important role in contemporary cosmological scenarios \cite{linde}. Vigourously 
evaporating black holes may produce interesting consequences in the 
inflationary epoch if our considerations regarding supersymmetry breakdown are valid. We 
hope to report on some of these issues elsewhere.

\end{document}